# Single-shot refractive index slice imaging using spectrally multiplexed optical transfer function reshaping

CHUNGHA LEE,[1,2] HERVE HUGONNET,[1,2] JUYEON PARK,[1,2] MAHN JAE LEE,[2,3] WEISUN PARK,[1,2] AND YONGKEUN PARK[1,2,4]

[1]*Department of Physics, Korea Advanced Institute of Science and Technology (KAIST), Daejeon, Republic of Korea*
[2]*KAIST Institute for Health Science and Technology, KAIST, Daejeon, Republic of Korea*
[3]*Graduate School of Medical Science and Engineering, KAIST, Daejeon, Republic of Korea*
[4]*Tomocube Inc., Daejeon, Republic of Korea*
*\*yk.park@kaist.ac.kr*

**Abstract:** The refractive index (RI) of cells and tissues is crucial in pathophysiology as a noninvasive and quantitative imaging contrast. Although its measurements have been demonstrated using three-dimensional quantitative phase imaging methods, these methods often require bulky interferometric setups or multiple measurements, which limits the measurement sensitivity and speed. Here, we present a single-shot RI imaging method that visualizes the RI of the in-focus region of a sample. By exploiting spectral multiplexing and optical transfer function engineering, three color-coded intensity images of a sample with three optimized illuminations were simultaneously obtained in a single-shot measurement. The measured intensity images were then deconvoluted to obtain the RI image of the in-focus slice of the sample. As a proof of concept, a setup was built using Fresnel lenses and a liquid-crystal display. For validation purposes, we measured microspheres of known RI and cross-validated the results with simulated results. Various static and highly dynamic biological cells were imaged to demonstrate that the proposed method can conduct single-shot RI slice imaging of biological samples with subcellular resolution.

## 1. Introduction

The refractive index (RI) can serve as a label-free imaging marker for biological cells and tissues [1]. Owing to its noninvasive and quantitative nature, RI has been widely used to image and analyze various samples, from cell biology to preclinical applications, including immunology [2-5], cell biology [6-10], hematology [11], regenerative medicine [12], and neuroscience [13, 14]. Recently, complex biological systems have also been investigated, including organoids [15], micro-vasculatures [16], *C.elegans* [17], embryos [18], and tissues [19]. Three-dimensional (3D) quantitative phase imaging (QPI) methods have been utilized to measure the RI distribution of samples [1, 20]. In the 3D QPI, multiple 2D optical phase images of a sample are measured to reconstruct the 3D RI distribution. 3D QPI methods exploit laser interferometry to record multiple holograms of the transmitted light fields [21-24]. However, an interferometric microscope with a coherent light source is complicated and also susceptible to environmental and coherent noise, which generates speckle noise [25].

To address these issues, various 3D QPI methods that employ incoherent light sources have been demonstrated [26-34]. Incoherent 3D QPI methods require the measurement of multiple intensity images to reconstruct the RI distribution of a sample under various imaging conditions. Representative examples include defocus imaging [26, 32], introducing diffractive masks [27], modulating image fields [18, 28], stitching multiple low-resolution images [29], and varying the illumination conditions [30, 31, 33]. Several QPI methods have recently demonstrated single-shot volumetric imaging of biospecimens by introducing microlens arrays [35] or pinhole arrays [36], controlling the polarization states of scattered light [37], and exploiting spectral multiplexing [38]. Nevertheless, the aforementioned single-shot QPI methods require complicated instrumentation [37, 38], are based on interferometry [35, 37, 38], or have a limited spatial resolution [36].

Here, we present a single-shot QPI method to directly obtain an RI slice image at a specific axial plane of a 3D sample using optical transfer function (OTF) reshaping and spectral multiplexing (Fig. 1(a)). While differential phase contrast has been widely applied to incoherent QPI methods to obtain phase information of a thin sample [33, 39, 40], the proposed method allows direct RI reconstruction of a thick sample at a specific axial plane. Three optimized illumination patterns were spectrally multiplexed into one color pattern to reshape the OTF of an imaging system. The transmitted color intensity images of the sample illuminated with this optimized color illumination pattern were recorded using a color camera. The RI distribution of the sample was retrieved from the deconvolution of the measured color intensity image. For demonstration, a compact and cost-effective setup was built using Fresnel lenses and a liquid crystal display (LCD), and various samples were measured and analyzed. We also present measurements of biological cell dynamics.

## 2. Methods

*2.1 Experimental Setup*

To demonstrate single-shot RI slice imaging, an optical setup with a color illumination unit was built (Fig. 1(b)). A white light-emitting diode (LED) (Luminus Device Inc., CBT-140-WDH-L16-QB220) was used as the illumination source. The beam was collimated in a Fourier plane using a Fresnel lens (Edmund Optics Inc., #13-458). The intensity distribution was then modulated in the Fourier plane using an LCD (Innolux Inc., AT070TNA2), followed by another Fresnel lens (Edmund Optics Inc., #13-458) which projects the illumination onto a sample. Note that white-light illumination is spectrally filtered via its propagation through the LCD, on which an 8-bit RGB image, composed of three optimized input source patterns for red (R), green (G), and blue (B) channels, is displayed. The beam transmitted through a sample was imaged on a color imaging sensor (XIMEA, MD120CU-SY) using an objective lens with a numerical aperture (NA) of 0.95 (Olympus Inc., UPlanXApo 40X), an achromatic tube lens, and an additional 4-$f$ telecentric imaging system with two achromatic lenses (L1, $f$ = 100 mm; L2, $f$ = 400 mm). The effective size of the pupil was cropped to an NA of 0.67 using an iris to avoid aberration at the pupil edge because of the imperfect collimation of spatiotemporally incoherent illumination. The effective NA of the setup can be further increased to enhance the spatial resolution by increasing the illumination field of view.

A representative single-shot color image of a HEK293T cell is shown in Fig. 1(c). The color image was then decomposed into three intensity images for the R, G, and B channels (Fig. 1(d)). A color correction approach was used to correct the spectral crosstalk of the color imaging sensor [41]. The subtraction of the decomposed intensity images from one another produced a pair of intensity images of the HEK293T cell, which correspond to the raw data for the two optimized illumination patterns of deconvolution phase microscopy (Fig. 1(e)). The RI slice image of the cell was obtained via deconvolution of the raw data using the engineered OTF (Fig. 1(f)).

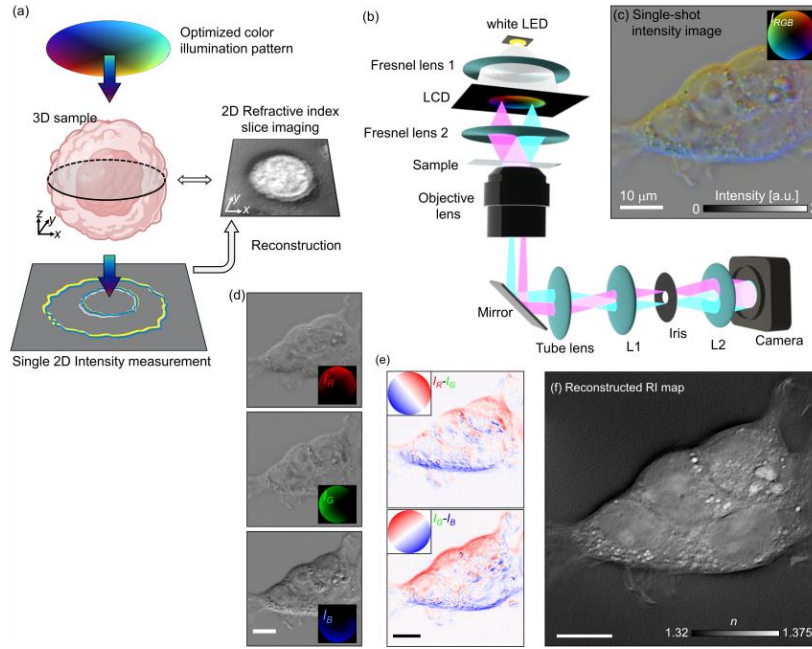

**Fig. 1** Single-shot RI slice imaging. (a) Concept of the proposed method. (b) Experimental setup. L1–L2, achromatic lenses; LCD, liquid-crystal display. (c) Single-shot color image of the fixed HEK293T cell. (d) Three intensity images for R, G, and B channels obtained from the decomposition of (c). Insets: source patterns displayed on the LCD. (e) Raw data for deconvolution phase microscopy obtained from the subtractions of (d) with one another. Insets: optimized illumination patterns of deconvolution phase microscopy. (f) The RI slice image of the HEK293T cell obtained via the deconvolution of (e).

## 2.2 Principles of Deconvolution Phase Microscopy

Deconvolution phase microscopy illuminates a sample using incoherent light with different angular distributions. Then, the phase information can be directly extracted from the intensity measurements via deconvolution of the measured intensity images [42, 43] (See Supplement 1). In order to perform 2D phase deconvolution of a 3D sample to obtain phase information, the sample is often assumed to be thin and in focus so that the sample spectrum is constant in the axial direction [39, 40]. In the proposed method, two illumination patterns, $\rho_{1,2}(k_{i,x},k_{i,y})$, were optimized for direct RI reconstruction of a sample at the focus plane. Instead of assuming the constant spectrum of the sample scattering potential in the axial direction, here the OTF is engineered to be constant in the axial direction so that the 3D deconvolution can be reduced to a 2D deconvolution [44]. This will enable the RI distribution to be obtained without axial scanning of the sample. Further, to image both the positive and negative parts of the illumination patterns, a total of four images must be measured [44]. Color multiplexing approach has been previously exploited in deconvolution phase microscopy to achieve single-shot acquisition of different complementary illumination patterns [46-51]. Likewise, we reduced the number of required images to three and assigned each pattern to a color channel, enabling single-shot imaging of the RI distribution at a specific axial plane. The three source patterns are defined as follows:

$$\rho_{RGB} = \begin{cases} \rho_R = \rho_1(\rho_1 > 0) + \rho_2(\rho_2 > 0) \\ \rho_G = \rho_2(\rho_2 > 0) - \rho_1(\rho_1 < 0) \\ \rho_B = -\rho_1(\rho_1 < 0) - \rho_2(\rho_2 < 0) \end{cases}, \quad (1)$$

which comprises only positive values to be displayed on the LCD as an 8-bit RGB image. The three intensity images corresponding to the three source patterns were then reduced into a pair of intensity images by simple subtraction from each other, corresponding to the illumination patterns, $\rho_1 = \rho_R - \rho_G$, $\rho_2 = \rho_G - \rho_B$. The RI distribution of the sample was obtained by solving Eq. (1), deconvolving the pair of intensity images using the pseudoinverse of the OTF matrix [45].

Figure 2 shows the optimized illumination patterns and corresponding OTF. The experimentally measured illumination patterns were consistent with the numerically calculated patterns (Fig. 2(b)). For precise illumination modulation, numerically calculated source patterns were manually calibrated by measuring the source patterns (Fig. 2(a), see Fig. S1 in Supplement 1). The OTF calculated using the experimentally measured illumination patterns exhibited uniform OTF density in the radial direction and constant OTF values in the axial direction (Fig. 2(c)).

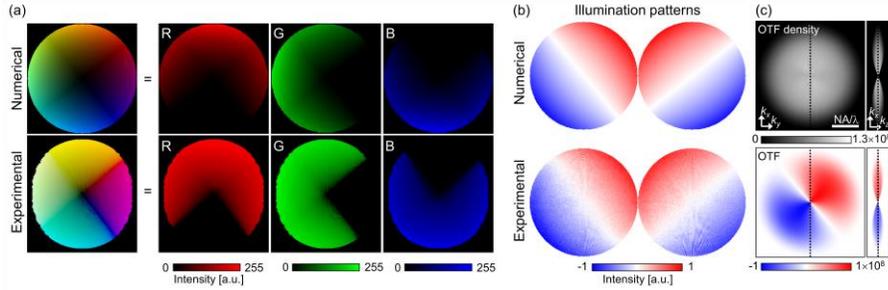

**Fig. 2** Optimized illumination patterns and the engineered OTF. (a) Numerical (top) and experimentally calibrated (bottom) source patterns, respectively. (b) Illumination patterns obtained numerically (top) and experimentally (bottom). (c) OTF density (top) and the OTF (bottom) calculated from the experimentally measured illumination patterns. λ is the peak wavelength for the B channel, λ = 452 nm, and NA is the effective NA of the optical setup.

*2.3 Sample Preparation*

For the microsphere samples, 8-μm-diameter polymethylmethacrylate (PMMA) microspheres ($n$ = 1.49 at 553 nm) were mounted in an ultraviolet curing resin (Norland Products Inc., NOA 148). HEK293T (ATCC, CRL-1573) cells were maintained in Dulbecco's modified Eagle's medium (DMEM; ATCC, 30-2002) supplemented with 10% fetal bovine serum (Thermo Fisher Scientific Inc.) and 1% (v/v) penicillin/streptomycin (Thermo Fisher Scientific Inc.) at 37°C in a 5% $CO_2$ incubator. Peripheral blood mononuclear cells (PBMCs) and polymorphonuclear leukocytes (PMNs) were collected from peripheral blood in an EDTA tube by density gradient centrifugation. Briefly, 10 mL of whole blood was diluted 1:1 with commercialized buffer, autoMACS® Rinsing Solution (Miltenyi Biotec Inc.) containing bovine serum albumin (Miltenyi Biotec Inc.). The diluted blood (20 mL) was then gently layered on top of 20 mL of Histopaque®-1077 and Histopaque®-1119 (Sigma-Aldrich Co.) and centrifuged continuously at 800 g for 30 min at 22°C. Mononuclear and polymorphonuclear layers were carefully collected and washed twice. The collected PBMCs and PMNs were resuspended in RPMI-1640 (Thermo Fisher Scientific Inc.) supplemented with 10% (v/v) fetal bovine serum (Thermo Fisher Scientific Inc.) and 1% (v/v) penicillin/streptomycin (Sigma-Aldrich Co.). All samples were loaded into imaging dishes (TomoDish, Tomocube Inc.) at a density of $1 \times 10^{-4}$ and $1 \times 10^{-6}$ cells/ml for HEK293T cells and blood cells, respectively. Peripheral blood was obtained from a healthy volunteer with the approval of the Internal Review Board (IRB) of KAIST (approval. No. KH2017-004). All procedures followed the Helsinki Declaration of 2000, and informed consent was obtained from all the participants.

## 3. Results

*3.1 Validation of the Proposed Method using Microspheres*

To validate the proposed method, we imaged an 8-μm-diameter PMMA microsphere ($n$ = 1.49 at a center wavelength of 553 nm) in an ultraviolet curing resin ($n_m$ = 1.48). Fig. 3(a) shows the experimental and ideal 3D RI distributions of the microsphere. In the numerical simulation, the 3D RI tomogram of the microsphere was calculated by a convolution between the Fourier-transformed 3D microsphere object and the OTF. The experimentally measured 3D RI tomogram matches the numerically calculated 3D RI tomogram. The line profiles in Fig. 3(b) confirm that the experimental results are consistent with the numerical results. Despite the consistent results, the sample RI values in both results were underestimated relative to the ground-truth RI value ($n = 1.49$). This underestimation results from the missing cone problem [52], where the spatial frequencies of the sample information are beyond the NA of an inaccessible lens. The missing-cone problem can be alleviated by applying a regularization algorithm [52]. Using the measured 3D RI tomogram, experimental resolutions were quantified as 0.82 and 3.16 μm in the lateral and axial directions, respectively. These measured spatial resolutions are consistent with the expected values [53]. In addition, the spatial and temporal noises of the system were measured as 6.81 and $9.28 \times 10^{-4}$, respectively (Supplement 1). The measured spatial sensitivity of our single-shot method is comparable to the reported sensitivity of the previous method, which requires four frames for RI slice reconstruction [44]

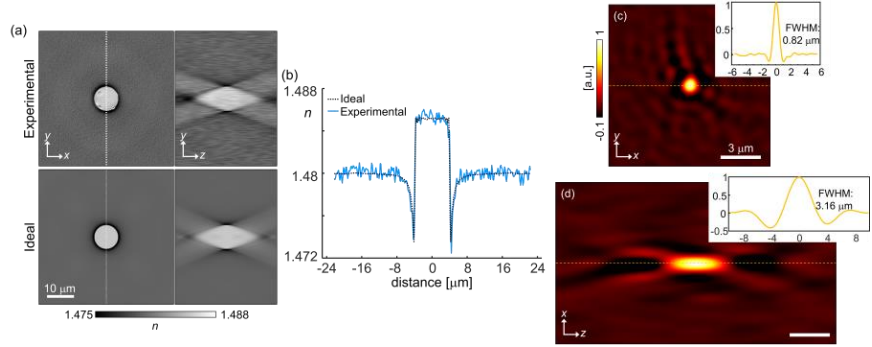

**Fig. 1** Validation of the proposed method. (a) Experimentally (top) and numerically obtained (bottom) 3D RI tomograms of an 8-μm-diameter PMMA microsphere, respectively. (b) Line profiles of the 3D RI tomograms corresponding to the dashed lines in (a). (c) and (d) *XY*- and *XZ*-slice images of experimentally measured point spread function (PSF). Insets: line plots of the PSF of the dashed lines in (c) and (d), respectively, in which experimental resolutions were quantified as the full-width-at-half-maximum (FWHM) of the PSF.

## 3.2 Single-shot RI Slice Imaging of Biological Cells

Single-shot RI slice imaging of the biological cells was performed using the proposed method. Representative RI slice images of both fixed HEK293T cells and live PBMCs are shown in Fig. 4. The RI images of fixed HEK293T cells in Figs. 4(a) and 4(b) display both the thin edge of the cell exhibiting membrane ruffling (i) [54] and the nucleolus covered by the nuclear envelope (ii) [55]. To further emphasize the subcellular resolution of the proposed method, we measured live PBMCs, which were three times smaller than HEK293T cells. In Figs. 4(c) and 4(d), the nucleus (iii) and microvilli (iv) of the PBMCs were clearly imaged, suggesting the broad applicability of the proposed method to pathophysiological studies involving subcellular RI analysis of live cells [56-58].

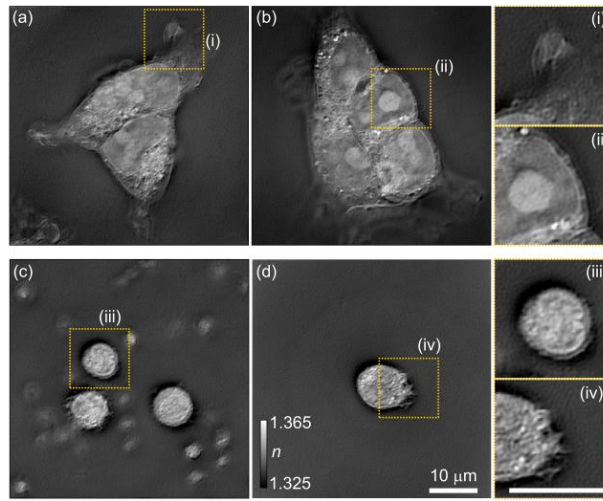

**Fig. 2** Single-shot RI slice images of fixed HEK293T cells and live PBMCs. (a)–(d) Reconstructed RI slice images of (a), (b) the HEK293T cells, and (c), (d) PBMCs. (i)–(iv) Magnified views of the dashed boxes in (a)–(d), which display (i) the thin edge of the cell and (ii) nucleolus inside the nuclear envelope of the HEK293T cells, and (iii) the nucleus and (iv) microvilli of the PBMCs.

By exploiting the single-shot RI slice imaging capability of the proposed method, dynamic measurements of migrating cells were demonstrated (Fig. 5, see Visualization 1 and Visualization 2). For each time-lapse measurement, 200 single-shot color images of live PMN cells were recorded at 0.82−1.22 fps. From the time-series data, a series of RI slice images of moving cells were obtained (Fig. 5). Indeed, the cells actively migrated with the dynamic extension of their pseudopods while crossing each other (Fig. 5(a)). Another PMN cell line also exhibited fast migration with dynamic extension and contraction of the cell membrane (Fig. 5(b)). These results confirmed that the proposed method could be utilized for time-lapse analyses of various cellular activities, such as cell growth [59], migration [60], and cell–cell communication [4].

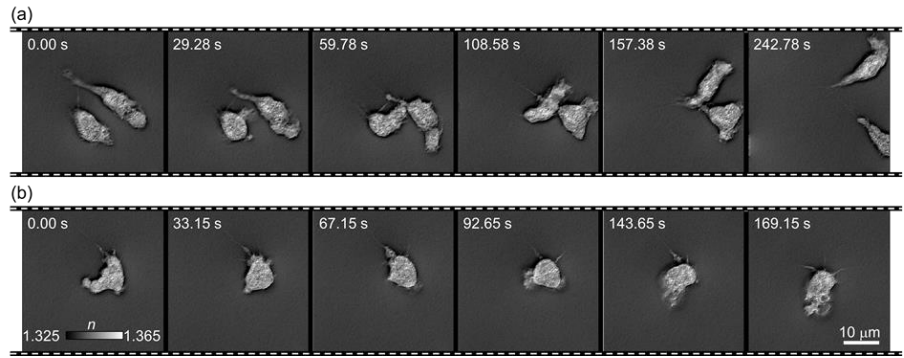

**Fig. 3** Time-series RI slice images of live PMN cells. (a) and (b) A series of RI slice images of (a) two PMN cells (see Visualization 1) and (b) a single PMN cell (see Visualization 2).

### 3.3 3D RI imaging of Biological Cells

Volumetric RI imaging of biological cells was also demonstrated by the axial scanning of the cells (Fig. 6). For each 3D RI tomogram, 40 single-shot color images were measured with a z-scan step set to 585 nm on a motorized z-stage (Trinamic Inc., TMCM-6212). Fig. 6 presents 3D rendering and *XY*-slice images of fixed HEK293T cells and live PBMCs. The 3D rendering of the measured 3D RI tomograms was performed using Tomostudio (Tomocube Inc.) (Figs. 6(a) and 6(c)). In the *XY*-slice images of the HEK293T cell shown in Fig. 6(b), different inner-cell components were observed at different focal planes, such as the cell membrane (i), nucleoli (ii), and high-RI components such as lipid droplets (iii) [61, 62]. The RI slice images of the 3D RI tomogram of PBMCs and platelets were also visualized at different focal planes (Fig. 6(d)). It was observed that the platelets mostly lay on a single plane (iv) because they were attached to the glass substrate of the imaging dish. Likewise, the PBMCs were placed on a cover glass. Nevertheless, volumetric 3D RI imaging of PBMCs provided additional information on the 3D spatial distributions of the inner cellular components (v and vi).

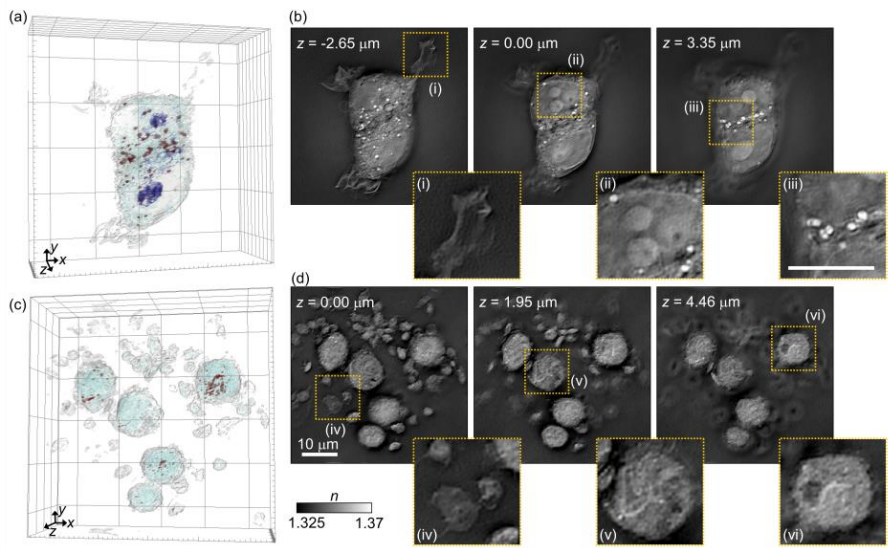

**Fig. 4** 3D RI tomograms of the fixed HEK293T cell and live PMMA cells. (a) and (b) 3D rendering and *XY*-slice images of the fixed HEK293T cell, respectively. (i)–(iii) Magnified views of the dashed boxes in (b), which display (i) the cell membrane, (ii) nucleoli inside the nuclear envelope, and (iii) high-RI components inside the cell. (c) and (d) 3D rendering and *XY*-slice images of live PBMC cells, respectively. (iv)–(vi) Magnified views of the dashed boxes in (d), which display (iv) the platelets, and (v), (vi) PBMCs of different morphologies.

## 4. Conclusions

We present a single-shot RI slice-imaging method that uses spectral multiplexing and OTF engineering. The proposed method was validated by comparing the imaging results on the microsphere samples with the theoretical results. Various cells, including HEK293T cells, PBMCs, and PMN cells, were imaged to demonstrate the applicability of our method to biological samples and its ability to resolve various subcellular features. Time-lapse imaging and volumetric RI imaging of cells were performed. The spatial and temporal sensitivities of our method were measured as 6.81 and $9.28 \times 10^{-4}$, respectively.

Compared with existing QPI techniques, the proposed method allows single-shot RI imaging of transparent 3D objects while being compact and cost-effective. The use of a temporally incoherent light source ensures high RI measurement sensitivity. The optical setup was designed to be compact and cost-effective by employing Fresnel lenses and an LCD in the illumination component. The 2-mm working distance of the Fresnel lens not only improves the user experience of the proposed method, but also suggests the potential for application to complex 3D biological systems, such as organoids, on-chip devices and 3D cell culture models [63-65]. For example, the proposed method may be used for time-lapse monitoring of developing organoids [66] or micro-vessels cultured in 3D microenvironment [16, 67]. To resolve more highly dynamic events in life, one may employ a color camera with high speed and signal-to-noise ratio for faster imaging, as the temporal resolution of proposed method is mainly limited by low signal-to-noise ratio and imaging speed of the color camera.

To further improve the image quality, more studies are needed to advance the RI reconstruction algorithm for spectrally multiplexed data. The intensity contrast of the image for the blue channel was remarkably stronger than those of the red and green channels. To address this challenge, we evaluated a simple normalization linear to the wavelength (see Fig. S2 in the Supplement 1), which resulted in no significant improvements, suggesting that a more advanced algorithm is required. Nevertheless, severe multiplexing artifacts were not found in the reconstructed RI images presented here. Considering the single-shot RI imaging capability of the proposed method, we envision that the proposed method could have wide-ranging applications in cell pathophysiology, immunology, and developmental biology, where visualization of living systems or high-throughput RI imaging of highly dynamic biospecimens is required.

**Funding.** This work was supported by Tomocube Inc., National Research Foundation of Korea (2015R1A3A2066550, 2022M3H4A1A02074314), and Institute of Information & communications Technology Planning & Evaluation (IITP; 2021-0-00745) grant funded by the Korea government (MSIT).

**Disclosures.** Y.P. has financial interests in Tomocube Inc., a company that commercializes optical diffraction tomography and quantitative phase imaging instruments and is one of the sponsors of the work.

**Data availability.** Full-resolution images presented in this study are available on request. Correspondence and requests for materials should be addressed to Y.P.

**Supplemental document.** See Supplement 1 for supporting content.

# Supplement Text

## Principles of Deconvolution Phase Microscopy

Deconvolution phase microscopy illuminates a sample using incoherent light with different angular distributions. Then, the phase information can be directly extracted from the intensity measurements via deconvolution of the measured intensity images [42, 43]. The intensity variation between the light transmitted through the sample and the illumination, $S = (I_{out}-I_{in})/I_{in}$ is described as

$$S = H_A V_{imag} + H_p V_{real}, \qquad (2)$$

where $V$ is the sample's scattering potential, $V = (1/4\pi) \cdot k^2 \left[ n^2(x,y,z)/n_m^2 - 1 \right] = V_{real} + i \cdot V_{imag}$, $k = |\mathbf{k}| = (k_x, k_y, k_z)$, $n$ is the 3D RI distribution of a sample, $n_m$ is the medium RI, and ~ indicates the Fourier transform [44]. Importantly, $H_A$ and $H_p$ are the amplitude and phase OTFs of an imaging system, respectively, which directly determine the imaging quality of the deconvolution phase microscopy. In particular, when *Köhler* illumination is implemented, the OTFs are expressed as the incoherent sum of the transmitted intensities for various spatial frequencies:

$$\begin{cases} H_A(\mathbf{k}) = \int \rho(k_{i,x}, k_{i,y}) H_A(\mathbf{k}; \mathbf{k_{in}}) dk_{i,x} dk_{i,y} \\ H_P(\mathbf{k}) = \int \rho(k_{i,x}, k_{i,y}) H_P(\mathbf{k}; \mathbf{k_{in}}) dk_{i,x} dk_{i,y} \end{cases}, \qquad (3)$$

where $H_{A,P}(\mathbf{k}; \mathbf{k_{in}}) = \mp 1/(k_z + k_{i,z}) \delta\left(k_z + k_{i,z} - \sqrt{k^2 - k_x'^2 - k_y'^2}\right) \pm i/(k_z + k_{i,z}) \delta\left(k_z + k_{i,z} - \sqrt{k^2 - k_x'^2 - k_y'^2}\right)$, $\mathbf{k'} = \mathbf{k} + \mathbf{k_{in}}$, $\mathbf{k_{in}}$ is the wavevector of the incident light, and $\rho(k_{i,x}, k_{i,y})$ is the intensity distribution or illumination pattern controlled in the Fourier plane [45].

**Calibration of the source patterns**

For accurate RI reconstruction in deconvolution phase microscopy, precise modulation of the illumination is required. With this aim, numerically calculated source patterns were manually calibrated by measuring the source patterns (Fig. S1). First, the nonlinear intensity responses of the LCD were calibrated by measuring the pupil images with white images displayed on the LCD (Fig. S1(a)). A total of 256 pupil images were measured with pixel values of 0-255 displayed on the LCD, from which the nonlinearity in the raw LCD responses was quantified for the three channels (Fig. S1(a), top). The nonlinear intensity responses, $I$, were then linearized using the fit function, $I(x) = a_1 \cdot \exp[-\{(x-b_1)/c_1\}^2]$, where $a_1$, $b_1$, and $c_1$ are the estimated coefficients and $x$ is the pixel value of 0-255 (Fig. S1(a), bottom). R-squared values of the fit functions are $> 0.999$ for three intensity responses. Second, image aberrations including image elongation, Barrel distortion, and intensity non-uniformity in pupil images, were detected by measuring the pupil mask with a grid pattern (Fig. S1(b)). The elongation of the pupil mask was corrected manually by a factor of 0.94. The Barrel distortion, due to the use of high-NA flat Fresnel lenses, was corrected via coordinate transformation of the target pupil image, $(r,\theta) \rightarrow (r', \theta')$, where $r' = (sin(tan^{-1}(\theta))$. The intensity non-uniformity in the measured pupil image was corrected by multiplying the target pupil image by the inverse of the measured pupil image. For precise modulation of the optimized illuminations, the aberration-corrected source patterns were further iteratively updated by measuring the $n$-th patterns and adding the difference between the numerically obtained ground truth and measured $n$-th patterns, $\Delta I_n(k_x, k_y) = I_{GT}(k_x, k_y) - I_{\rho_{n-th}}(k_x, k_y)$, to the $(n+1)$-th patterns.

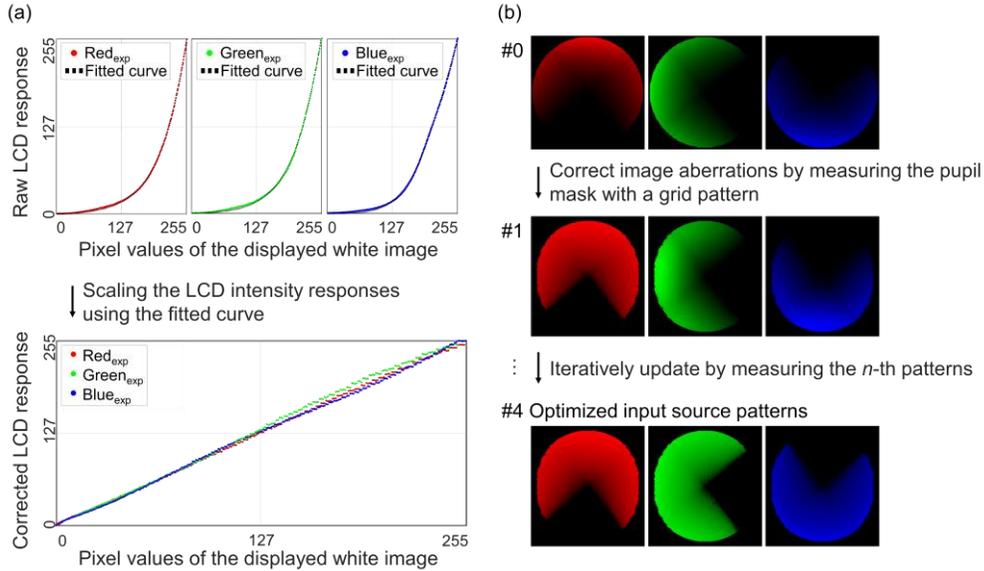

**Fig. S5** Calibration of the source patterns. (a) Linearization of the LCD responses. (b) Correction of image aberrations and iterative update of the aberration-corrected source patterns.

**Evaluation on a simple approach for wavelength normalization**

To address the issue of higher intensity contrasts at shorter wavelengths, we evaluated a simple normalization approach using the three peak wavelengths of the incoherent illumination, 453, 513, and 596 nm for R, G, and B channels (Fig. S2). Specifically, the intensity contrasts were normalized by multiplying the three intensity images by the corresponding weights of 453/513, 513/513, and 598/513. Compared to the results without normalization, no significant differences were found in the normlized raw data as well as the reconstructed RI image obtained from deconvolution of the normalized raw data (Figs. S2(a) and S2(b)). For this reason, we did not apply this normalization approach throughout the present work.

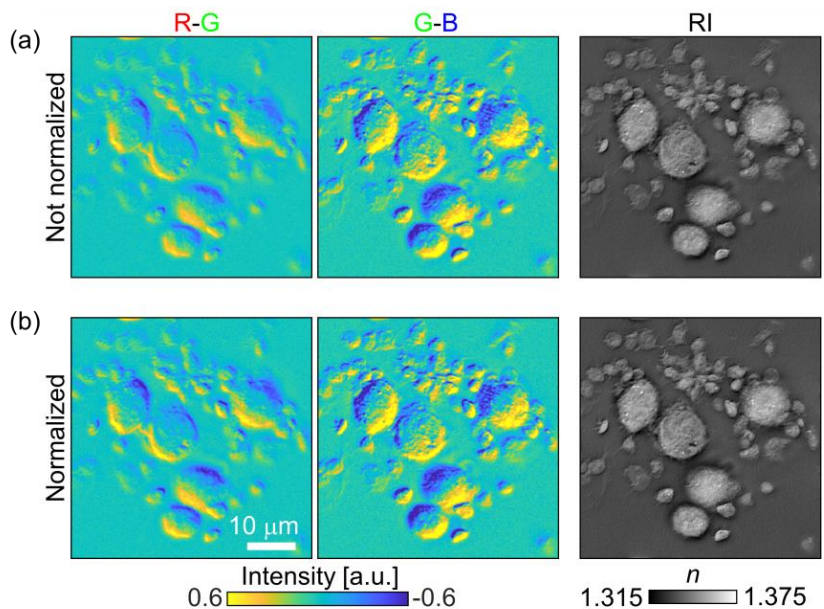

**Fig. S2** Evaluation on a simple normalization approach. (a) and (b) Raw data for deconvolution phase microscopy and the reconstructed RI image of PBMCs (a) without and (b) with applying the normalization approach, where no significant differences were found.

**Sensitivity Characterization**

To evaluate the effect of system noise on image quality, a fixed PBMC was measured using the proposed method. Fifty frames of static PBMC were measured, from which spatial and temporal noise were characterized using the reconstructed RI slice images in the background region of interest (ROI, 100 × 100 pixels) (Fig. S3(a), the inset at the top left). While spatial noise was calculated as the standard deviation of the single RI slice image, temporal noise was quantified as the spatial average of the temporal standard deviation for a series of 50 RI slice images. Thus, the spatial and temporal noises of the system were measured as 6.81 and $9.28 \times 10^{-4}$, respectively. The measured spatial sensitivity of our single-shot method is comparable to the reported sensitivity of the previous method, which requires four frames for RI slice reconstruction [44]. No apparent improvements were observed in resolving the details of the PBMC with averaging frames (Fig. S3(a)). Nevertheless, the spatial sensitivity was significantly reduced with averaging frames of $6.81 \times 10^{-4}$ without averaging frames, and 5.02, 3.89, 3.15, 2.84, 2.69, and $2.63 \times 10^{-4}$, averaging 2, 4, 10, 20, and 50 frames, respectively (Figs. S3(b) and S3(c)). The two-fold noise reduction in the spatial sensitivity when averaging four frames suggests that the proposed method is mainly limited by shot noise. When averaging more frames, the noise level was not reduced as much as expected because of the presence of fixed-pattern noise, which could have been induced by free-floating particles in the cell culture medium or dust in the optical system.

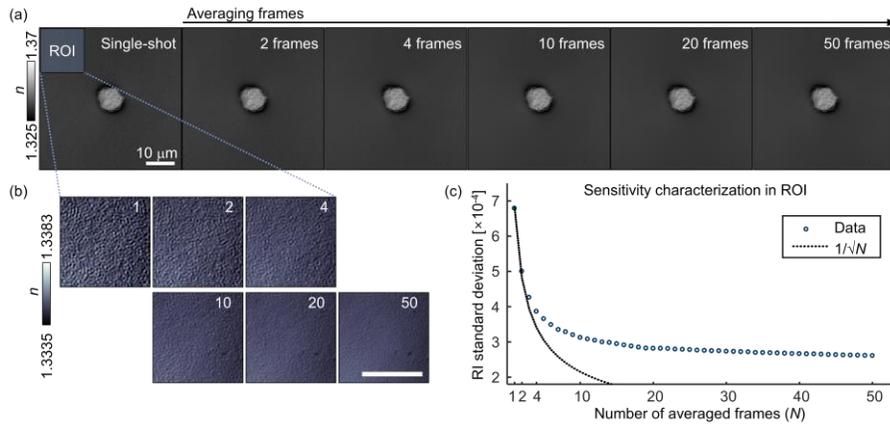

**Fig. S3** Sensitivity characterization with averaging frames. (a) RI slice images of the fixed PBMC obtained without averaging frames, and averaging 2, 4, 10, 20, and 50 frames, respectively. (b) Magnified views of a background region of interest (ROI) in (a). (c) Quantification of the spatial sensitivity with averaging frames.